# Exploring Multimodal AI Reasoning for Meteorological Forecasting from Skew-T Diagrams

ChangJae Lee[a,b], Heecheol Yang[c], Jonghak Choi[d]

[a] *Forecast Bureau, Korea Meteorological Administration, Seoul, Korea*

[b] *M.S. in Data Science Program, The University of Texas at Austin, Austin, TX, USA*

[c] *Distributed Networks and Computing Laboratory, Division of Computer Convergence, Chungnam National University, Daejeon, Korea.*

[d] *AIPIM, Seoul, Korea*

*Corresponding author*: Heecheol Yang, hcyang@cnu.ac.kr

# ABSTRACT

Forecasting from atmospheric soundings is a fundamental task in operational meteorology, often requiring structured visual reasoning over Skew-T log-P diagrams by human forecasters. While recent advances in Vision-Language Models (VLMs) have shown promise in other scientific domains, their application to meteorological diagram interpretation remains largely unexplored.

In this study, we present a lightweight AI assistant that interprets Skew-T diagrams using a small language model (LM) and a small VLM fine-tuned to emulate human forecasters. Using a curriculum learning framework, we first train the models to identify key atmospheric features from diagrams through visual question answering, followed by chain-of-thought reasoning tasks that estimate precipitation probability based on the derived visual groundings. Model inputs include either textual summaries or generated Skew-T diagrams derived from operational Numerical Weather Prediction (NWP) forecasts, paired with three-hour precipitation observations from South Korea's Auto Weather Stations network.

Evaluation results demonstrate that the fine-tuned VLM achieves skill comparable to an operational NWP model, despite relying solely on static atmospheric profiles. Ablation studies reveal that visual grounding and reasoning supervision are critical for performance, while attention map analysis confirms that the model learns to focus on relevant meteorological features.

These findings highlight the potential of compact, interpretable multimodal models to support weather forecasting tasks. The approach offers a computationally efficient alternative to large-scale systems, and future work could extend it to more complex applications.

# SIGNIFICANCE STATEMENT

In many fields, large language models and multimodal models are now widely adopted for their capacity to perform human-like reasoning. Their ability to integrate textual and visual information makes them especially effective for tasks that demand structured interpretation and informed decision-making. However, despite their broad adoption, applications in meteorology, such as interpreting weather charts, specialized diagrams, or remote-sensing imagery, remain limited. This study addresses this gap by showing that a compact, fine-tuned vision-language model can emulate human forecasters in analyzing Skew-T log-P diagrams and predicting precipitation probability. The proposed approach offers a computationally efficient and interpretable alternative to large-scale models, highlighting the potential of domain-specific AI systems to enhance operational forecasting and decision support in weather-related sectors.

## 1. Introduction

Large Language Models (LLMs) and Multimodal Models are increasingly being applied across diverse domains, including biomedicine, physics, chemistry, and economics. Their strengths in automation and efficiency contribute to increased productivity, and furthermore, their capacity for human-like reasoning enables them to support both experts and non-experts in complex tasks (Zhang et al., 2025). Motivated by these advances, we present an early-stage weather-AI assistant designed to reason like a human forecaster, with a focus on interpreting Skew-T log-P diagrams (Hess, 1959), one of the fundamental tools for analyzing and diagnosing atmospheric soundings.

Human forecasters typically interpret these diagrams through a step-by-step process: first diagnosing individual features, and then drawing an integrated conclusion based on their combined interpretation. We adopt this structured approach in our Vision-Language Model (VLM) by training these sequential steps.

Rather than employing LLMs, we fine-tuned a Small Language Model (SLM) and a small VLM that can both effectively achieve our goals while offering energy efficiency. Such small models, considering models below 10 billion parameters, are more adaptable for fine-tuning and can achieve strong performance on domain-specific tasks (Belcak et al., 2025).

To evaluate our multimodal weather-AI assistant, we designed two sets of tasks aligned with the step-by-step reasoning process conducted by weather forecasters: The first assesses the ability for visual-grounding, such as diagnosing atmospheric instability, wind profiles, and humidity conditions. The second estimates precipitation probability by classifying it into four categories, low, moderate, high, and very high, based on the analysis of a Skew-T log-P diagram.

As AI continues to support analysis and decision making processes, the development of interpretable models tailored to domain-specific tasks has become increasingly important. Our work addresses this need by presenting an efficient, human-like reasoning system tailored for meteorological applications.

Code is available at https://github.com/hunter3789/VLM-Skew-T.

## 2. Methods and Data

*2.1. Task setting*

Our goal is to train both the Language Model (LM) and VLM to interpret atmospheric sounding profiles and estimate precipitation probability based on their analysis. To support this, we implemented an agent that has a data processing module that converts raw atmospheric sounding inputs into two formats: (1) a summarized textual representation and (2) a Skew-T log-P diagram, generated from a tool-calling step. These processed inputs are then provided to the LM and VLM to generate answers. Figure 1 illustrates the outline structure of our task.

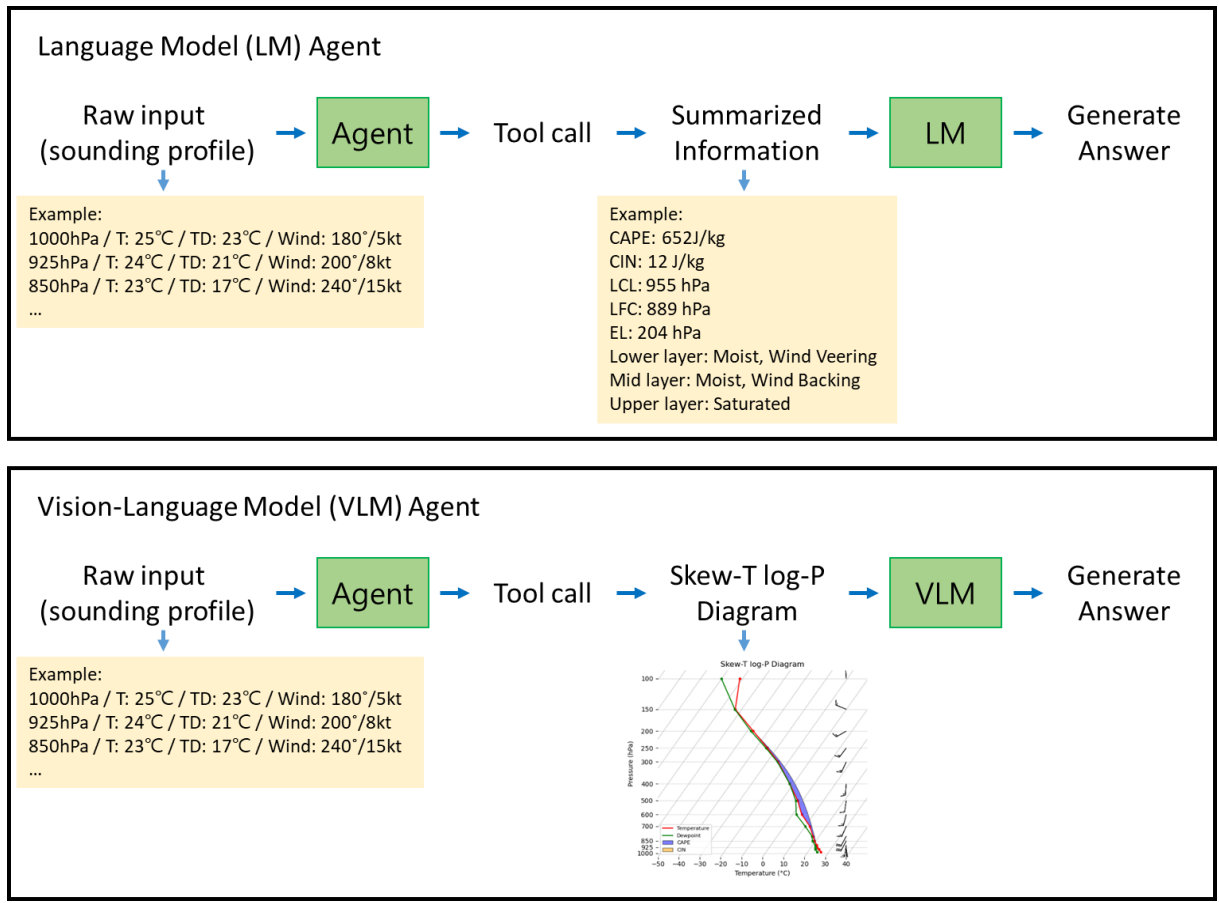

*Figure 1. Overview of the outline for the language model and vision-language model agents used to forecast precipitation probability from atmospheric sounding data.*

To achieve this goal, we fine-tuned a small Language Model (LM) and a small VLM using Low-Rank Adaptation (LoRA; Hu et al., 2021). Specifically, we used SmolLM2 (Allal et al., 2025) as our baseline LM and SmolVLM (Marafioti et al., 2025) as the baseline VLM. These models are from small family models which are designed to be computationally efficient, with parameters ranging from 250 million to 2 billion, while still achieving strong performance on targeted tasks.

*2.2. Training strategies*

In a multimodal setting, effective performance requires both visual grounding and reasoning based on visual analysis. To address this, we adopt a curriculum learning approach (Bengio et al., 2009), gradually increasing the complexity of the training tasks. First, we train the model using simple Visual Question Answering (VQA; Agrawal et al., 2016) samples focused on visual grounding. Then, we introduce more complex reasoning tasks using Chain-of-Thought (CoT; Wei et al., 2023) samples to guide the model through step-by-step inference.

Prior to training, we set a system prompt for the VLM to guide its interpretation of Skew-T log-P diagrams. The prompt anchors key features, including relevant terminology, legends, and basic interpretation tips, to help the model focus on critical aspects of the diagram. For structured analysis, we partitioned the atmospheric profile into three layers: the lower layer (1000-850 hPa), the mid-layer (850-500 hPa), and the upper layer (500-250 hPa). The system prompt used during both training and inference is shown in Table 1.

{"system": "You are a weather forecaster analyzing atmospheric soundings shown in a Skew-T log-P diagram.

The diagram uses a logarithmic vertical pressure axis (hPa), so pressure layers are not evenly spaced.

Use the following visual anchors:

- Lower layer (1000-850 hPa): This is located in the bottom quarter of the diagram, close to the surface. It represents the boundary layer where surface temperature, dew point, and CIN typically appear.

- Mid layer (850-500 hPa): Appears in the second quarter from the bottom of the plot. This region often contains most of the CAPE and developing updrafts.

- Upper layer (500-250 hPa): This is around the middle third of the diagram, despite covering less pressure range. This layer includes the top of convection (EL), cirrus clouds, and upper-level wind shear.

Use the following visual references:

- Red line: temperature profile

- Green line: dew point temperature

- Blue shaded area: CAPE (Convective Available Potential Energy)

- Yellow shaded area: CIN (Convective Inhibition)

> - Wind barbs: on the right-hand side, changing with height
>
> Key interpretation rules:
>
> - Where the red and green lines are close, the layer is moist; far apart implies dryness.
>
> - The LFC is the bottom of the blue area; the EL is the top of the blue area.
>
> - Clockwise turning wind barbs with height suggest veering (warm air advection); counterclockwise suggests backing."}

*Table 1. System prompt used during training and inference, including visual interpretation guidance for Skew-T log-P diagrams.*

In Stage I of the curriculum learning setup, we fine-tuned the model using VQA samples to guide its attention toward relevant regions of the diagram. These samples included questions targeting key atmospheric features such as humidity (classified as dry, moist, or saturated) and wind profile (veering or backing) for each vertical layer, and instability (strong, moderate, or weak), based on pre-defined diagnostic criteria. Further details of the VQA sample construction are provided in Table 2. Each training sample was answered with explanation.

| Types | Class 1 | Class 2 | Class 3 | Layers |
|---|---|---|---|---|
| Humidity | Dry | Moist | Saturated | Lower layer |
| | Average dew point depression > 3°C | Average dew point depression > 1°C | Average dew point depression <= 1°C | Mid layer Upper layer |
| Wind profile | Veering | Backing | - | Lower layer |
| | Rotation > 30° | Rotation < -30° | | Mid layer Upper layer |
| CAPE | Weak | Moderate | Strong | - |
| | CAPE < 500 J/Kg | CAPE < 1000 J/Kg | CAPE ≥ 1000 J/Kg | |
| CIN | Weak | Moderate | Strong | - |
| | CIN ≤ 75 J/Kg | CIN ≤ 150J/Kg | CIN > 150 J/Kg | |
| LFC | Low | Moderate | High | - |
| | LFC > 900 hPa | LFC > 700 hPa | LFC ≤ 700 hPa | |
| EL | Low | Moderate | High | - |
| | EL > 400 hPa | EL > 250 hPa | EL ≤ 250 hPa | |
| VQA Sample 1 | "Q": "Is the lower layer (1000-850 hPa, the bottom quarter of the diagram) dry, moist, or saturated? Please provide both explanation and answer.", "A": "The temperature (red) and dew point (green) lines are moderately close in the lower layer (1000-850 hPa, the bottom quarter of the diagram), suggesting fair humidity. Therefore, the lower layer is moist." | | | |
| VQA Sample 2 | "Q": "Is the CAPE in the diagram strong, moderate, or weak? Please provide both explanation and answer.", "A": "The blue CAPE area is present and spans a modest vertical range, which suggests some atmospheric instability, supportive of convection, but not extreme. Therefore, the CAPE is moderate." | | | |

*Table 2. Description of the VQA dataset construction, including classification criteria and sample questions.*

In Stage II training, we fine-tuned the model using reasoning samples focused on estimating precipitation probability based on the analysis of atmospheric sounding profiles. All training samples were formatted using a CoT structure to encourage efficient and structured reasoning. For the LM, training inputs were generated from textual summaries of the sounding data, including thermodynamic indices, humidity and wind conditions. In contrast, training data for the VLM incorporated visual grounding interpretations derived from Skew-T log-P diagrams. A rule-based approach was employed to generate prompts for both models by applying threshold-based criteria to key atmospheric variables. To introduce linguistic diversity, appropriate sentences for each condition were sampled from a predefined bag-of-phrases. After diagnosing the atmospheric profile, the model concluded each response with a precipitation probability for the next three hours categorized into four levels: low (no rain), moderate ($\geq 0.1$ mm/ 3 h), high ($\geq 1$ mm/ 3 h), and very high ($\geq 5$ mm/ 3h). Examples of these reasoning samples are provided in Table 3.

| LM training example | "System": "You are a weather forecaster analyzing atmospheric soundings shown in a Skew-T log-P diagram.", "User": "CAPE: 835 J/kg CIN: 52 J/kg LCL: 944 hPa LFC: 807 hPa EL: 149 hPa Lower layer: Moist Mid layer: Moist, Wind Backing Upper layer: Saturated, Wind Backing", "Assistant": "835 J/kg of CAPE suggests sufficient energy for thunderstorm development, given appropriate environmental triggers. A low LCL (~944 hPa) implies efficient cloud formation at the lower layer. The LFC at 807 hPa indicates parcels must traverse a neutral or stable layer before initiating buoyant ascent. An EL of 149 hPa is indicative of strong updraft potential and storm tops penetrating into the upper troposphere or lower stratosphere. The atmospheric column exhibits sufficient humidity to support rain-producing cloud layers. Based on the analysis, the probability of precipitation is high." |
|---|---|
| VLM training example | "System": omitted (see Table 1 for full prompt description),<br><br>"User": "Please describe the atmospheric profile based on the provided Skew-T log-P diagram. Reason carefully, and conclude with a precipitation probability category: Low, Moderate, High, or Very High",<br><br>"Image":<br>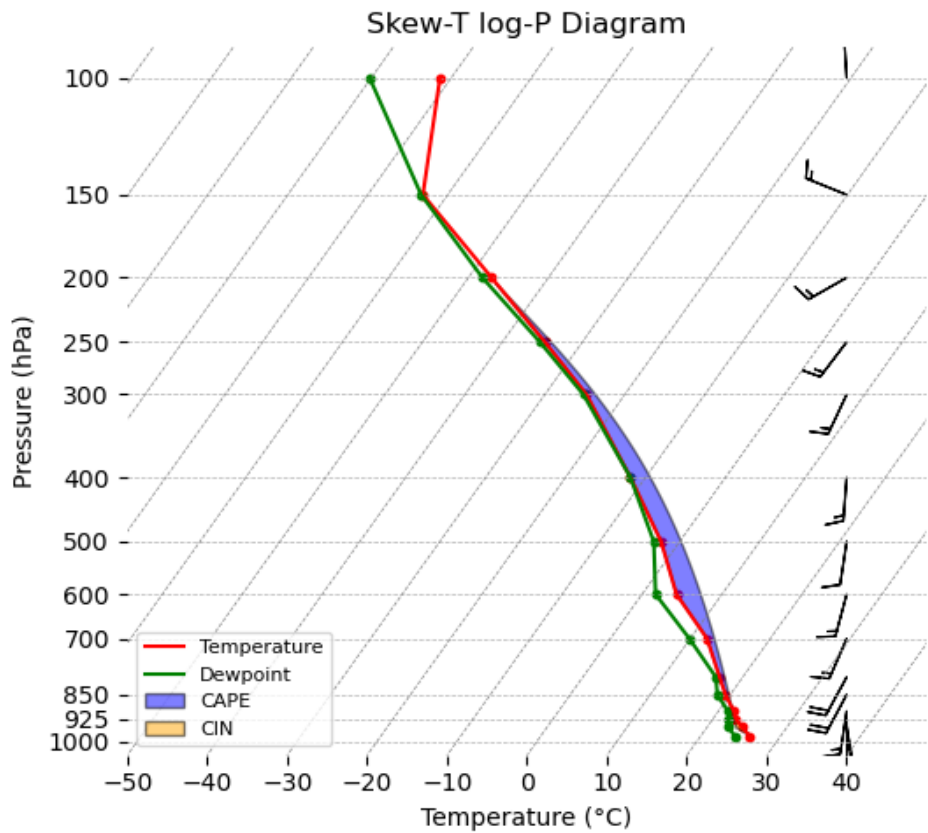,<br><br>"Assistant": "A moderately sized blue-shaded region (CAPE) implies moderate instability, which may support thunderstorm development if lifting is present.<br><br>The lower boundary of the blue region sits at a moderate height, requiring moderate vertical motion for convection to initiate.<br><br>The blue region extends high into the upper atmosphere, indicating the potential for very tall convective towers and vigorous updrafts. |

|   | Red and green lines lie close together from the surface through the mid and upper levels, indicating widespread moisture and favorable conditions for precipitation. |
|---|---|
|   | Based on the analysis, the probability of precipitation is very high." |

*Table 3. An example of reasoning samples used for LM and VLM training.*

*2.3. Visual-grounding*

To understand which visual tokens a VLM attends to when encountering keywords associated with diagram features, we visualized the model's self-attention maps conditioned on key phrases (Das et al., 2016). Figure 2 presents attention maps from the baseline model, while Figure 3 shows the corresponding maps after fine-tuning. Attention weight maps are obtained for each visual token in a $9 \times 9$ grid by averaging the attention weights across all word tokens in the key phrase and across all layers and heads of the transformer architecture (Vaswani et al., 2023). To visualize overall attention trends over visual tokens, a Gaussian filter with a standard deviation of 1 is applied to present smooth maps. The resulting attention values are then normalized before visualization to highlight which visual tokens are attended more to each phrase. Note that SmolVLM's image processor applies pixel shuffling (Shi et al., 2016), which increases the effective receptive field beyond the size of the corresponding visual token.

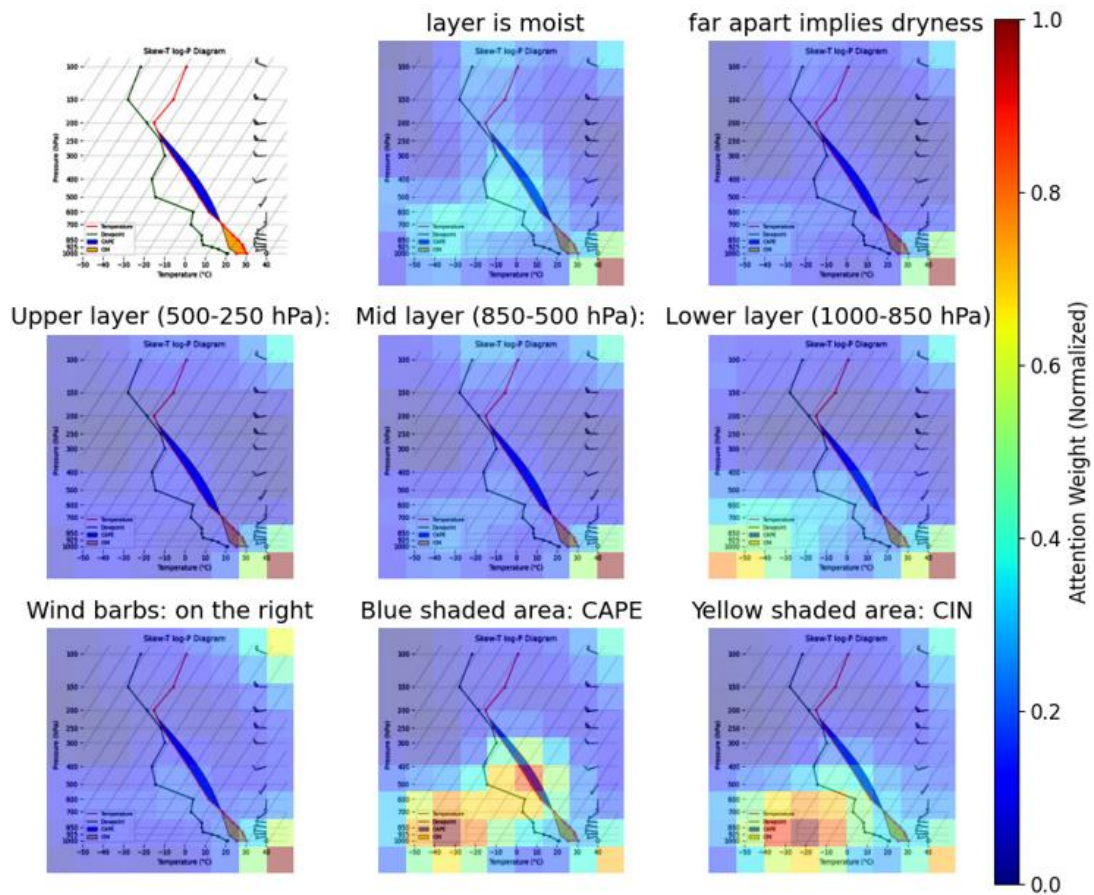

*Figure 2. Self-attention maps from the baseline VLM showing visual attention patterns for key phrases.*

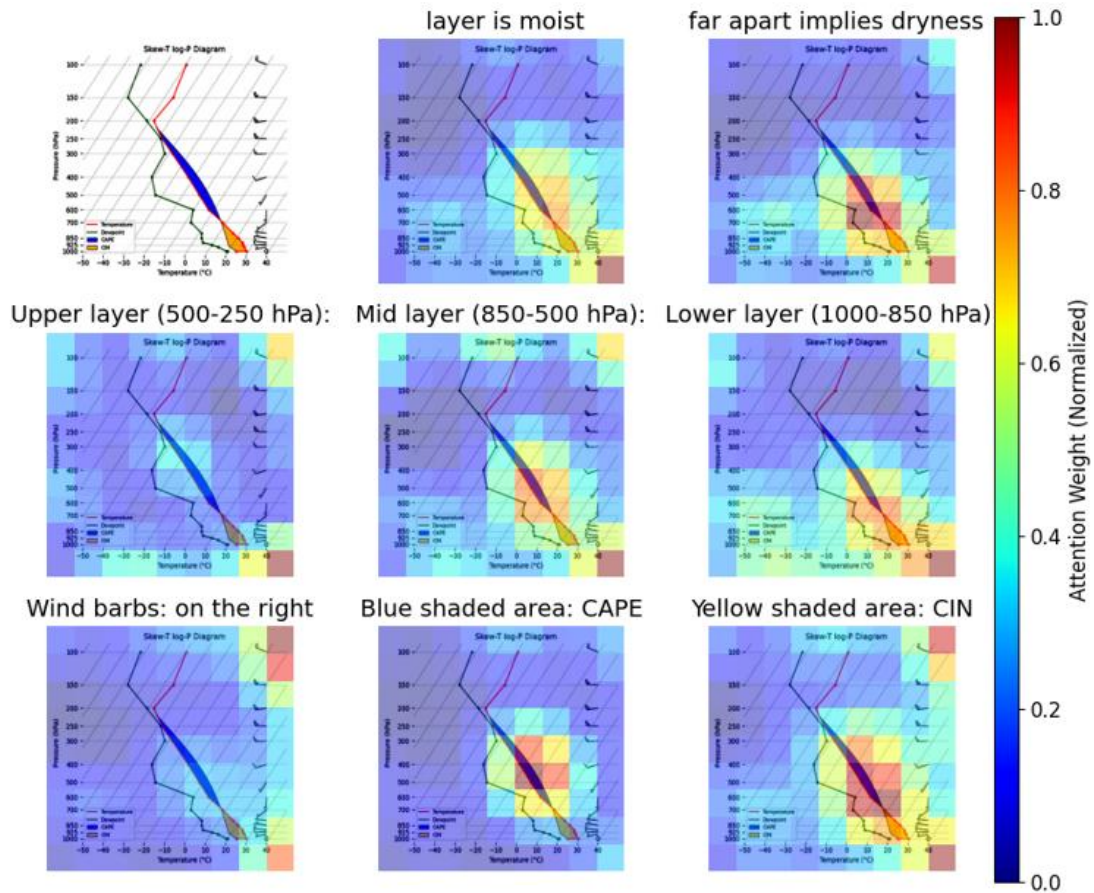

*Figure 3. Self-attention maps from the fine-tuned VLM showing improved visual alignment with key phrases.*

Figure 2 shows that the baseline model, SmolVLM-2.2B, exhibits limited attention to the image when no explicit visual anchors, such as specific colors or keywords, are present. However, because SmolVLM was pre-trained on diverse datasets involving chart and figure understanding, as well as optical character recognition (OCR), it tends to focus on diagram elements like legends or color-coded regions when the phrase contains matching keywords. For instance, in response to the prompt "Blue shaded area: CAPE", the model attends to the corresponding colored region and the legend.

In contrast, Figure 3 represents that the fine-tuned model, which is trained with VQA samples specific to Skew-T log-P diagrams, exhibits focused attention on the relevant parts of the diagram. For instance, when prompted with phrases related to humidity, the model attends to the vertical profiles of temperature and dewpoint temperature. Similarly, when given prompts about vertical layers or wind profiles, the model aligns its attention with the appropriate diagram regions: the upper layer corresponds to the middle third of the diagram, the mid-layer to the second-lowest quarter, the lower layer to the bottom quarter, and wind barbs to the right side of the diagram. Notably, after fine-tuning, the model no longer attends strongly to the static legend when interpreting CAPE or CIN, as it learns that the legend is not directly relevant for answering such questions.

*2.4. Training and validation datasets*

To generate the training dataset, we used European Centre for Medium-Range Weather Forecasts' (ECMWF) Integrated Forecasting System High-Resolution (IFS-HRES; Rasp et al., 2023) model output, focusing on summer seasons from 2021 to 2023. Data were sampled every three hours, using forecast lead times from 0 to 9 hours for the 00 and 12 UTC runs. Atmospheric sounding profiles were extracted for 10 representative cities in South Korea. Since our task is to reason about the rainfall possibility for the upcoming three-hour time window, these profiles were paired with the next three-hourly precipitation observations from Auto Weather Stations (AWS), using rain-gauge measurements as the ground truth. The combined data were then transformed into structured formats suitable for model training, as illustrated in Table 2 and 3.

The validation dataset followed the same format but was derived from the summer season (June, July, and August) of 2024, which was excluded from the training data to ensure an independent evaluation. Under this setup, the training dataset consists of 140,608 VQA samples and 21,810 reasoning samples, while the validation dataset contains 48,355 VQA samples and 7,350 reasoning samples.

*2.5. Training settings*

As part of our two-stage curriculum learning approach, we trained the model for 0.2 epochs on Stage I (VQA) and 4 epochs on Stage II (CoT reasoning), with a replay ratio (Portelas et al., 2020) of 0.1 for Stage I during Stage II training. The VQA dataset (140,608 samples in total) is larger than the CoT dataset (21,810 samples in total), as multiple questions are generated for each atmospheric profile, thus conducting fewer epochs for initial training. We applied LoRA to the baseline model, using the following hyperparameters: rank r = 16, scaling factor $\alpha$ = 64, dropout = 0.05-0.1, and learning rate = 1e-04.

*2.6. Evaluation strategies*

For evaluation, we fine-tuned models under various configurations to identify the most effective approach for mimicking human forecasters. Model (a) is a LM baseline trained on textual summaries of atmospheric profiles, while model (b) is a VLM baseline trained using Skew-T log-P diagram inputs. Models (c) and (d) share the same VLM architecture as model (b), but differ in training setup: model (c) is trained without Chain-of-Thought (CoT) reasoning supervision to assess the importance of CoT during training, and model (d) is trained with multimodal fusion inputs (image + text) to evaluate their usefulness. Models (e) and (f) are scaled-up versions of models (b) and (d), respectively, designed to examine the impact of model size. Finally, model (g) represents an operational Numerical Weather Prediction (NWP) system (IFS-HRES), used as a reference to benchmark the performance of VLMs against a state-of-the-art forecasting model.

|     | Model | Input | CoT-Reasoning | Notes |
| --- | --- | --- | --- | --- |
| (a) | SmolLM2-360M | Textual summary | ✓ Yes | Fine-tuned LM Baseline |
| (b) | SmolVLM-250M | Image only | ✓ Yes | Fine-tuned VLM Baseline |
| (c) | SmolVLM-250M | Image only | ✗ No | No reasoning supervision |
| (d) | SmolVLM-250M | Image + Textual summary | ✓ Yes | Tests usefulness of multimodal fusion input |
| (e) | SmolVLM-2.2B | Image only | ✓ Yes | Model scaling up (2.2B) |

| (f) | SmolVLM-2.2B | Image + Textual summary | ✓ Yes | Tests usefulness of multimodal fusion input on a larger model |
| (g) | IFS-HRES | - | - | Operational NWP |

*Table 5. Summary of model configurations used for benchmarking.*

We employed a range of evaluation metrics to assess both VQA performance and the model's ability to forecast precipitation probability. For question answering tasks, accuracy and F1-score (Bishop, 2006) were used to measure classification performance across various categories. Given the imbalanced distribution of precipitation events and the focus on identifying rainfall occurrences for each criteria, we used Frequency Bias (F.Bias), Critical Success Index (CSI), Probability of Detection (POD), and Success Ratio (SR) (Wilks, 1995) to evaluate the models' reasoning capabilities in forecasting tasks.

## 3. Evaluation results

Model verification was conducted using data from the summer of 2024, under the same configuration used during training. Consistent with the two-stage training framework, both VQA performance and reasoning capability were evaluated across all model variants.

*3.1. VQA results comparison*

We compared the VQA performance of fine-tuned SmolVLM models with 250 million and 2.2 billion parameters here. The VQA dataset including answers for validation was generated under the setup illustrated in Table 2. F1-scores were weighted according to the sample size of each class. The scores and sample sizes for each criteria are reported in Table 6.

For relatively straightforward tasks, such as diagnosing humidity, Convective Available Potential Energy (CAPE), and Equilibrium Level (EL), both models achieved strong performance, with both accuracy and weighted F1-scores exceeding 0.8.

In contrast, both models struggled with more complex or subtly defined tasks, including the diagnosis of wind profile, Convective Inhibition (CIN), and the Level of Free Convection (LFC). For example, CIN was categorized as weak ($\leq 75$ J/kg), moderate ($\leq 150$ J/kg), or strong (> 150 J/kg), making the classification task difficult due to the small differences between categories. Similarly, LFC was labeled as low (1000-900 hPa), moderate (900-700 hPa), or high ($\leq$ 700 hPa), posing a challenge since LFC values often occurs near 900 hPa. Diagnosing wind profiles as veering or backing is also complex, as it requires interpreting the directional rotation of wind barbs with respect to height, a skill that demands nuanced visual reasoning.

Nevertheless, the larger model (2.2B) demonstrated stronger performance on these challenging tasks compared to the smaller model (250M). With more attention heads and transformer layers, the 2.2B model exhibited improved visual grounding ability, as reflected in higher VQA scores across the complex diagnostic categories.

| Question types / sample sizes | | | SmolVLM-250M | | SmolVLM-2.2B | |
|---|---|---|---|---|---|---|
| | | | Accuracy | Weighted F1-score | Accuracy | Weighted F1-score |
| Humidity (three-layers) | | | 0.83 | 0.81 | 0.87 | 0.87 |
| Dry | Moist | Saturated | | | | |
| 17,574 | 3,096 | 1,116 | | | | |
| Wind profile (three-layers) | | | 0.51 | 0.49 | 0.65 | 0.65 |
| Veering | | Backing | | | | |
| 7,057 | | 7,228 | | | | |
| CAPE | | | 0.87 | 0.86 | 0.97 | 0.97 |
| Weak | Moderate | Strong | | | | |
| 1,039 | 817 | 1,665 | | | | |
| CIN | | | 0.55 | 0.46 | 0.70 | 0.70 |
| Weak | Moderate | Strong | | | | |
| 1,700 | 874 | 959 | | | | |
| LFC | | | 0.73 | 0.67 | 0.75 | 0.67 |

| Low | Moderate | High | | | | |
|---|---|---|---|---|---|---|
| 367 | 1,838 | 277 | | | | |
| EL | | | | | | |
| Low | Moderate | High | 0.92 | 0.92 | 0.99 | 0.99 |
| - | 127 | 2,355 | | | | |

*Table 6. VQA evaluation results across all question types for fine-tuned models with 250 million and 2.2 billion parameters. The leftmost column indicates the sample size for each category.*

*3.2. Reasoning benchmark*

We evaluated the reasoning performance of our models in predicting precipitation within the next three hours based on the interpretation of meteorological sounding data. Precipitation probability was categorized into three levels: moderate (0.1 mm/ 3 h), high (1 mm/ 3 h), and very high (5 mm/ 3 h). Performance scores for each category are reported in Table 7, 8, and 9. Note that the dataset is imbalanced: only 13.1% of cases exceeded the moderate threshold, 8.3% exceeded the high threshold, and 4.1% fell into the very high category.

For predicting whether it will rain or not, where probabilities of moderate or above, we observed that the conventional NWP model (IFS-HRES) tends to over-predict rainfall, whereas both the LM and VLM tend to under-predict it from Table 7. The under-prediction tendency observed in both the LM and VLM may be attributed to several factors, including class imbalance (with precipitation occurring in only ~15% of samples), conflicting meteorological signals, and limited contextual information. For example, similar sounding profiles can result in either rainfall or no rainfall depending on subtle atmospheric variations, making it challenging for the models to produce accurate predictions under such ambiguity.

Although the models tend to under-predict rainfall, the Critical Success Index (CSI), which accounts for both true positives and false positives, reported from the VLM is comparable to, or even exceeds, that of the conventional NWP model. In addition, several noteworthy findings emerge from the evaluation results:

Finding 1: A small VLM (250 million parameters, model (b)) can outperform a larger LM (360 million parameters, model (a)) when the image contains additional information that supports reasoning. Model (b) achieved a higher CSI score than model (a) by mitigating its under-prediction tendency through effective visual grounding.

Finding 2: CoT reasoning is crucial for teaching the VLM how to interpret sounding data. Model (c), which did not receive CoT-based training, performed the worst among all models, highlighting the importance of step-by-step reasoning in diagram analysis.

Finding 3: Model scaling is beneficial, but small models offer an efficient alternative. At the lower precipitation threshold (0.1 mm/ 3 h), scaling up from 250 million to 2.2 billion parameters did not result in a noticeable improvement in performance. However, at higher thresholds (as shown in Tables 7 and 8), the 2.2 billion parameter model (e) outperformed the smaller model (b). Despite this, the small model remains a competitive and computationally efficient option, achieving comparable results at a fraction of the resource cost.

Finding 4: Multimodal fusion inputs (image + text summary) can enhance performance when used with a model of sufficient capacity. For instance, the 2.2 billion parameter model (f) benefits from this fusion, showing improved scores. In contrast, the smaller model (d) (250 million parameters) shows degraded performance. This can be inferred from its VQA scores, where model (d) struggled with specific visual grounding tasks. The additional textual input may not align well with its weaker visual understanding. In contrast, model (f), which demonstrated stronger visual grounding capabilities with more attention heads and deeper transformer layers, is better equipped to leverage the complementary strengths of multimodal fusion.

|     | Model | Input | CoT-Reasoning | Moderate ($\geq$ 0.1 mm/ 3 h) | | | |
|     |       |       |               | Occurrence : ~ 13.1% | | | |
|     |       |       |               | F.BIAS | CSI | POD | SR |
| --- | --- | --- | --- | --- | --- | --- | --- |
| (a) | SmolLM2-360M | Text summary | ✓ Yes | 0.3 | 26.6% | 28.1% | 82.6% |
| (b) | SmolVLM-250M | Image only | ✓ Yes | 0.6 | 39.3% | 44.2% | 77.9% |

|     | Model        | Input        | CoT-Reasoning | F.BIAS | CSI   | POD   | SR    |
| --- | ------------ | ------------ | ------------- | ------ | ----- | ----- | ----- |
| (c) | SmolVLM-250M | Image only   | ✘ No          | 0.1    | 10.0% | 10.3% | 76.9% |
| (d) | SmolVLM-250M | Image + Text | ✓ Yes         | 0.4    | 32.6% | 35.0% | 82.8% |
| (e) | SmolVLM-2.2B | Image only   | ✓ Yes         | 0.6    | 38.5% | 43.8% | 75.9% |
| (f) | SmolVLM-2.2B | Image + Text | ✓ Yes         | 0.6    | 39.9% | 45.6% | 76.2% |
| (g) | IFS-HRES     | -            | -             | 2.5    | 33.8% | 87.1% | 35.5% |

*Table 7. Evaluation metrics for reasoning performance across models when predicting rainfall above 0.1 mm/ 3 h.*

For predicting rainfall above 1 mm/ 3 h, the overall performance trends remain similar to those observed at the lower precipitation threshold of 0.1 mm/ 3 h, as shown in Table 8. One notable result is that the VLM reduces its under-prediction tendency, with the frequency bias improving to approximately 0.8-0.9, closer to the ideal value of 1. This improvement may be attributed to that atmospheric sounding profiles associated with this threshold exhibit more coherent patterns, such as moist layers, veering wind profiles, and absolute unstable atmospheric environments, making them easier to interpret.

|     | Model        | Input        | CoT-Reasoning | High (≥ 1 mm/ 3 h) Occurrence : ~ 8.3% | | | |
| --- | ------------ | ------------ | ------------- | ------ | ----- | ----- | ----- |
|     |              |              |               | F.BIAS | CSI   | POD   | SR    |
| (a) | SmolLM2-360M | Text summary | ✓ Yes         | 0.5    | 30.6% | 35.8% | 67.9% |
| (b) | SmolVLM-250M | Image only   | ✓ Yes         | 0.9    | 38.7% | 52.5% | 59.5% |
| (c) | SmolVLM-250M | Image only   | ✘ No          | 0.2    | 9.7%  | 10.2% | 67.1% |
| (d) | SmolVLM-250M | Image + Text | ✓ Yes         | 0.6    | 35.9% | 43.2% | 68.0% |
| (e) | SmolVLM-2.2B | Image only   | ✓ Yes         | 0.9    | 38.4% | 52.7% | 58.7% |
| (f) | SmolVLM-2.2B | Image + Text | ✓ Yes         | 0.8    | 38.8% | 51.2% | 61.5% |
| (g) | IFS-HRES     | -            | -             | 1.6    | 41.2% | 75.7% | 47.4% |

*Table 8. Evaluation metrics for reasoning performance across models when predicting rainfall above 1 mm/ 3 h.*

Finally, for predicting rainfall above 5 mm/ 3 h, the performance trends remain similar, as shown in Table 9. At this higher threshold, all models tend to converge toward an ideal frequency bias close to 1. While the conventional NWP model achieves the best overall performance due to its integration of more comprehensive atmospheric information, the fine-tuned VLMs demonstrate a comparable ability to reason about precipitation based on a single snapshot of the atmospheric profile.

|     | Model | Input | CoT-Reasoning | Very high ($\geq$ 5 mm/ 3 h) Occurrence : ~ 4.1% | | | |
| --- | --- | --- | --- | --- | --- | --- | --- |
|     |       |       |               | F.BIAS | CSI | POD | SR |
| (a) | SmolLM2-360M | Text summary | ✓ Yes | 1.1 | 28.6% | 45.8% | 43.2% |
| (b) | SmolVLM-250M | Image only | ✓ Yes | 0.9 | 28.7% | 43.1% | 46.1% |
| (c) | SmolVLM-250M | Image only | ✗ No | 1.0 | 24.2% | 39.5% | 38.4% |
| (d) | SmolVLM-250M | Image + Text | ✓ Yes | 0.9 | 26.1% | 38.5% | 44.7% |
| (e) | SmolVLM-2.2B | Image only | ✓ Yes | 1.1 | 30.8% | 50.5% | 44.2% |
| (f) | SmolVLM-2.2B | Image + Text | ✓ Yes | 1.1 | 31.0% | 49.5% | 45.3% |
| (g) | IFS-HRES | - | - | 1.1 | 34.3% | 54.5% | 48.1% |

*Table 9. Evaluation metrics for reasoning performance across models when predicting rainfall above 5 mm/ 3 h.*

## 4. Discussion

From this work, we demonstrated that a fine-tuned small VLM can effectively mimic human forecasters and assist in weather forecasting tasks. The evaluation results highlight the potential of VLM in interpreting meteorological diagrams and reasoning about rainfall probability. Although the VLM exhibits an under-predicting tendency towards light rain (0.1

mm/ 3 h), the model's performance underscores its potential as a complementary tool for supporting weather forecasting, exhibiting competitive capabilities compared to conventional NWP models. While this study focused on summer seasons in the South Korea region, future work can explore the generalizability of the approach across different seasons and geographic regions to further validate the robustness of VLM-based forecasting systems.

Beyond the quantitative results, AI-based assistants offer additional advantages over human forecasters. (1) They can generate consistent and objective outputs, free from recency bias or fatigue (Murdock, 1962). (2) They can scale effortlessly by producing multiple responses in parallel, enabling rapid dissemination or ensemble-style interpretation.

Building on this foundation, future work can extend the approach to more complex tasks, such as interpreting weather charts, satellite imagery, or radar data. These advanced applications may involve the integration of multiple visual inputs and temporal sequences, introducing new challenges for model design and training. In addition to fine-tuning, incorporating techniques such as Retrieval-Augmented Generation (RAG; Lewis et al., 2021) could further enhance the model's capacity to utilize external contextual information. Overall, these directions offer exciting opportunities to further bridge AI and meteorological expertise.

*Acknowledgments*

This manuscript was prepared with the assistance of ChatGPT (OpenAI, 2025), which was used for language refinement and grammar correction. The authors take full responsibility for the content.

*Data Availability Statement.*

The code used in this study is available at https://github.com/hunter3789/VLM-Skew-T. The training and validation datasets used in this publication are accessible at https://osf.io/4n3uh/files/osfstorage.